\documentclass[ superscriptaddress, 
                notitlepage, 
                aps,
                prb,
                twocolumn,
                reprint] 
                {revtex4-1}
\usepackage{hyperref}
\usepackage{color,soul}
\usepackage{amsmath}
\usepackage{amssymb}
\usepackage[caption=false]{subfig}
\usepackage{graphicx}

\usepackage[normalem]{ulem}

\usepackage[colorinlistoftodos,prependcaption,textsize=tiny]{todonotes}

\newcommand{\rcut}{\ensuremath{r_{\text{cut}}}}
\newcommand{\cmt}
{
    Condensed Matter Theory Group,
    Paul Scherrer Institute,
    5232 Villigen PSI,
    Switzerland
}
\newcommand{\mesosysethz}
{
    Laboratory for Mesoscopic Systems,
    Department of Materials,
    ETH Zurich,
    8093 Zurich,
    Switzerland
}
\newcommand{\mesosyspsi}
{
    Laboratory for Multiscale Materials Experiments,
    Paul Scherrer Institute,
    5232 Villigen PSI,
    Switzerland
}
\date{\today}

\begin{document}


\title{
    Phase diagram of dipolar-coupled XY moments on disordered square lattices
}
\author{Dominik \surname{Schildknecht}}
\email{dominik.schildknecht@psi.ch}
\affiliation{\cmt}
\affiliation{\mesosysethz}
\affiliation{\mesosyspsi}
\author{Laura J. \surname{Heyderman}}
\affiliation{\mesosysethz}
\affiliation{\mesosyspsi}
\author{Peter M. \surname{Derlet}}
\affiliation{\cmt}
\begin{abstract}
    The effects of dilution disorder and random-displacement disorder are
    analyzed for dipolar-coupled magnetic moments confined in a plane, which
    were originally placed on the square lattice.
    In order to distinguish the
    different phases, new order parameters are derived and parallel tempering
    Monte Carlo simulations are performed for a truncated dipolar
    Hamiltonian to
    obtain the phase
    diagrams for both types of disorder. We find that both dilution disorder
    and random-displacement
    disorder give similar phase diagrams, namely disorder at small enough
    temperatures favors a so-called microvortex phase. This can be understood
    in terms of the flux closure present in dipolar-coupled systems.
\end{abstract}
\maketitle
\section{Introduction}
In frustrated magnetic systems, there is a variety of interesting phenomena
including the
inhibition of long-range order~\cite{Wannier1950}, highly
degenerate ground states~\cite{Pauling1935}, incommensurate
phases~\cite{Elliott1961,Fisher1980,Selke1988}, spin glass
physics~\cite{Edwards1975,Sherrington1975} and emergent rules on local
fluctuations~\cite{Ramirez1999,Castelnovo2008}. In many of these systems,
including the pyrochlores~\cite{Harris1997,Ramirez1999}, dipolar contributions
are
important. Moreover, the dipolar interaction
itself may be understood in terms of frustration. Here the ferromagnetic and
antiferromagnetic components compete, resulting in its anisotropic behavior.

In recent years, artificial spin systems, manufactured by assembling
single-domain nanoscale
magnets have been investigated~\cite{Heyderman2013,Nisoli2013,Marrows2016}.
These nanomagnets interact purely via magnetostatic coupling that, to lowest
order,
can be described by dipolar coupling only. In many of these artificial spin
systems, the nanomagnets have Ising-like degrees of
freedom~\cite{Wang2006,Heyderman2013,Farhan2013,Cumings2014,
Anghinolfi2015a,Sendetskyi2016}. In addition, a modification of the interaction
energies was recently demonstrated by combining Ising-like nanomagnets with
nanomagnets featuring continuous in-plane moments placed at the
vertices~\cite{Ostman2017}.

Systems entirely built out of dipolar-coupled moments that rotate freely in the
plane
are predicted to exhibit interesting physics such as continuously degenerate
ground states~\cite{Belobrov1983} and order-by-disorder
mechanisms~\cite{Prakash1990}.
Their experimental investigation is, however, still in its
infancy~\cite{Heyderman2004,Arnalds2014,Velten2017,Leo2018}.
Such a system will henceforth be denoted as a dXY system,
where the XY is in analogy to the XY model and the d refers to the dipolar
coupling.

Without any assumptions about the geometry of a dXY system, the only
symmetry supported by the Hamiltonian is time reversal. If the moments are
placed on a regular lattice, the symmetry group of the Hamiltonian is
enhanced by the point group of the lattice as a result of the anisotropy of
the dipolar interaction. Therefore, different geometries will give rise to
additional phases and universality classes for the transitions involved.

    If the dXY system is placed on the square lattice, the system is known
    to have a continuously-degenerate ground state, despite the symmetry group
    of the Hamiltonian being finite rather than continuous~\cite{Belobrov1983}.
    In previous work, the so called \textit{order-by-disorder} transition was
    demonstrated~\cite{Prakash1990}. Here finite temperature leads to
    an effective
    selection of certain states of the ground-state manifold due to different
    spin-wave stiffnesses along certain directions that follow the fourfold
    symmetry of the square lattice. This results in a low-temperature
    long-range ordered striped phase. A similar selection effect is seen with
    the introduction of disorder in the form of vacancies. Here a long-range
    ordered microvortex state emerges, that also respects the finite symmetry
    of the Hamiltonian~\cite{Prakash1990}.

For the non-disordered dXY system on the square lattice, the resulting phase
transition to a high-temperature paramagnetic regime has been studied
numerically, revealing either an Ising~\cite{Baek2011} or an
XYh4~\cite{DeBell1997,Carbognani2000,Fernandez2007} universality class
transition. The critical exponents obtained by numerical investigations lie
within the numerical error at the values expected for the Ising model.
But, since the XYh4 has a marginal operator, which means that the critical
exponents can be tuned to the critical
exponents of the Ising universality class in one of the limiting
cases~\cite{Jose1977a}, it is not clear
if the dXY system on the square lattice saturates
this limit and therefore belongs to the Ising universality class, or is just
close to saturation and is therefore only properly described by an XYh4
universality class. Consequently, numerical investigations
of this transition are to a certain degree inconclusive.

It can, however, conclusively be argued that the value of $h_4$ is
large~\cite{DeBell1997,Carbognani2000,Fernandez2007}, such that the system has a
strong effective anisotropy, that follows the fourfold anisotropy of the
square lattice. This drives the dXY system on the square lattice away from the
Berezinskii-Kosterlitz-Thouless
transition~\cite{Berezinskii1971,Kosterlitz1973}
towards a clear second order phase transition.

The \textit{dilution-disordered system}, where vacancies are introduced, was
previously studied using a temperature-sweep Monte Carlo approach and an
observable, which consisted of fourth powers of the spin
components~\cite{DeBell1997,Patchedjiev2005,LeBlanc2006}. Various values of
dilution were examined and well-converged results were obtained up to a
dilution
rate of approximately $6\%$. In addition, the results qualitatively agreed with
the
predictions from the spin-wave analysis~\cite{Prakash1990}.

The dXY system on the square lattice with \textit{random
displacement} of the sites was studied using a parallel tempering
approach~\cite{Alonso2011}. Here the spin glass overlap
observable was considered and it was concluded that no
spin glass phase is observed even for the highest amounts of disorder.
In addition, fully random placement of dipolar-coupled XY spins, as well as
random-displacement disorder applied to the square lattice, was studied by
means of a saddle-point analysis~\cite{Pastor2008}. Here, it was demonstrated
that a spatial localization of magnetic excitations occurs in systems with
strong disorder. To the best of our knowledge, however, no phase diagram has
been determined for a random-displacement disordered dXY system on the square
lattice.

In this paper, the full phase diagrams for both the dilution disordered as well
as the random-displacement disordered case are obtained numerically. Both
diagrams display a pocket at low temperature and moderate disorder where the
microvortex phase dominates. Furthermore, there is a striped phase region for
smaller disorder and higher temperature. Starting from either phase, the
paramagnetic regime is obtained if either temperature or disorder are increased
sufficiently. The structure of the phase diagram can be understood by
considering the magnetic flux closure present in dipolar systems. If the full
symmetry of the square lattice is present, the flux closure can occur
globally and the striped phase will dominate over the microvortex phase due to
a smaller spin-wave
stiffness. If the point group symmetries are broken by introduction of
disorder, magnetic flux closure will occur locally and the microvortex phase
will dominate at low temperatures.

The remainder of the paper is organized as follows. The model and the
order parameters are introduced in
section~\ref{sec:derivation_order_parameter}.
The methods are specified in section~\ref{sec:mc} and numerical data are
reported for the non-disordered case in section~\ref{sec:non-disordered}. Our
results for the dilution-disordered dXY system are presented in
section~\ref{sec:dilution} and for random displacement in
section~\ref{sec:random_displacement}. A Binder cumulant analysis is introduced
and consequently applied to the data in order to give a system-size
independent phase diagram. The limitations and the applicability of the order
parameters are then discussed in section~\ref{sec:applicability}, where it is
shown that, for the disorder range dealt with in this paper, the order
parameters are still well defined.
Finally, similarities between the phase diagrams for the two
types of disorder are highlighted in section~\ref{sec:conclusion} and
a possible interpretation of these similarities in terms of magnetic
flux closure is provided.
\section{Model \& Order Parameters}
\label{sec:derivation_order_parameter}
The (classical) Hamiltonian of the dXY system is given by
\begin{align}
    H=\frac{D}{2}\sum_{i\neq j}\frac{p_ip_j}{r_{ij}^3}
    \left[ \vec S_i\cdot\vec S_j-
        3\left( \vec S_i\cdot\hat r_{ij} \right)
        \left( \vec S_j\cdot\hat r_{ij} \right)
    \right],
    \label{eq:hamiltonian}
\end{align}
where the spins, as well as their positions, are confined to the $xy$-plane.
$D$ denotes the dipolar-interaction strength and without loss of generality is
set to~$1$. The dilution parameters $p_i$ are either $1$ or $0$, and are 0
if the $i$th moment is removed and 1 otherwise.
In the non-disordered system all $p_i$ are $1$.
The difference vector between the positions at the sites $i$ and $j$ is denoted
by $\vec r_{ij}$, its length by $r_{ij}$ and the normalization of this
vector to unit length is denoted as $\hat r_{ij}$.
For a non-disordered system, all sites lie on a regular square lattice in the
$xy$-plane, and the nearest-neighbor distance is set to~$1$. For the
introduction of random displacements,
the position of each site is randomly displaced in the $xy$-plane according to
a Gaussian distribution.

    For the remainder of the paper, a cutoff radius will be applied to the
    evaluation of Eq.~\eqref{eq:hamiltonian} in order to speed up the
    calculations.
    Therefore instead of a summation of all sites $i\neq j$, we will only
    consider contributions of sites with
    $1\leq|\vec r_{ij}|\leq\rcut$, where
    $\rcut$ is the cutoff radius. The cutoff chosen for the simulations in
    section~\ref{sec:mc} was $\rcut=2$, which included the 12 closest lattice
    sites. In what follows, we will abbreviate the studied system with tdXY
    for \emph{t}runcated \emph{d}ipolar \emph{XY}.
    The rather small value for \rcut~was chosen, since we expect that the
    qualitative features in the phase diagram will already be captured
    correctly, while the Hamiltonian can still be evaluated quickly so that
    extensive simulations
    can be performed. Note, however, that a larger value of $\rcut$ would
    result in
    more frustration, so that quantitative features such as the critical
    temperature $T_c$ are expected
    to decrease with larger $\rcut$.

\begin{figure}
    \subfloat[][]{
        \includegraphics[width=.5\columnwidth]{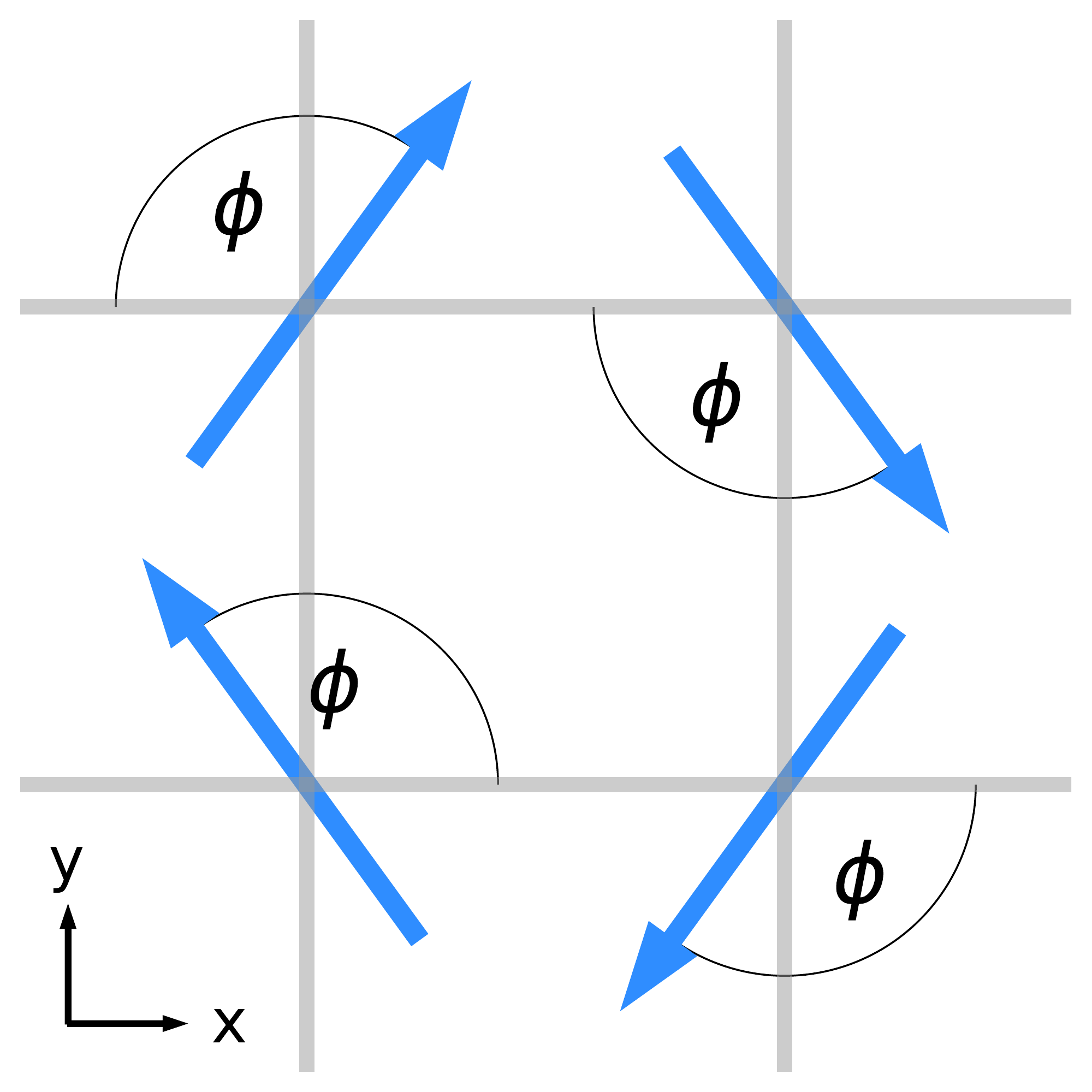}
        \label{fig:gs}
    }
    \subfloat[][]{
        \includegraphics[width=.5\columnwidth]{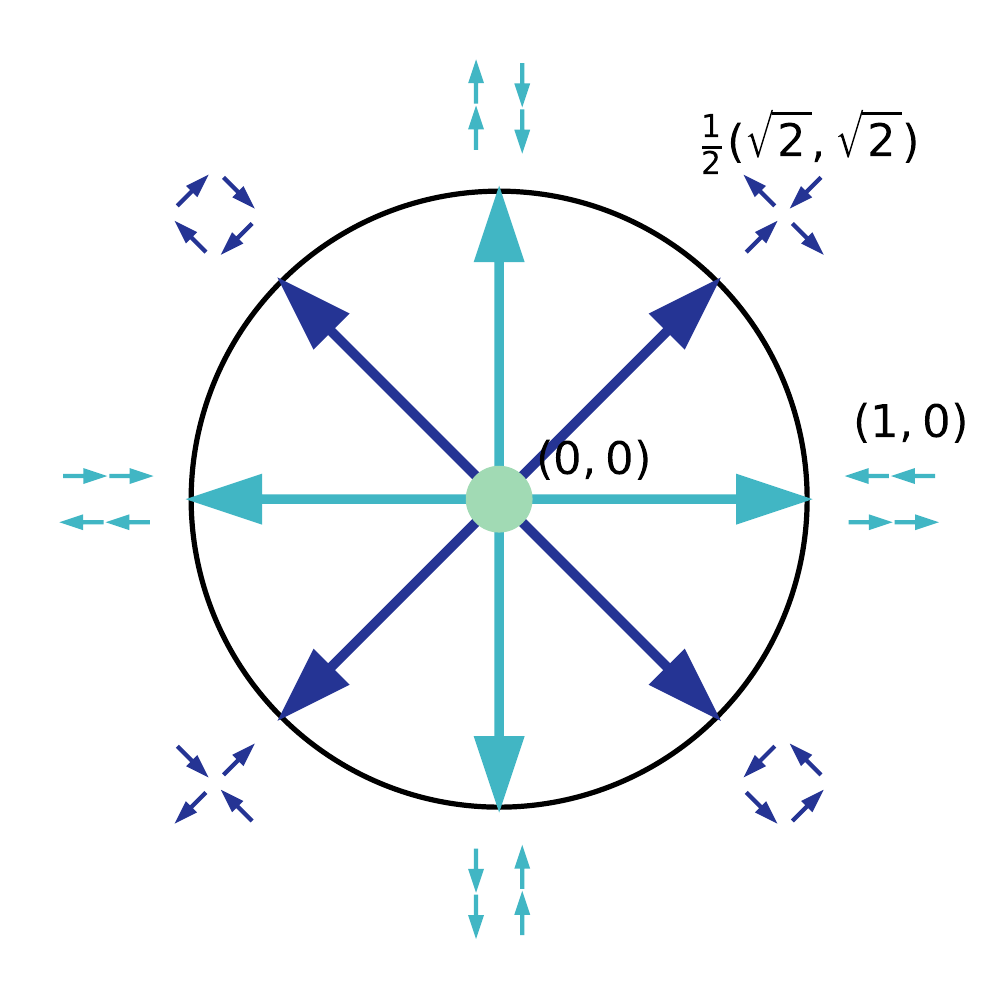}
        \label{fig:order-parameter-interpretation}
    }
    \caption{(color online)
        \textbf{(a)} The degenerate ground state of the square-lattice dXY
        system is defined within a two-by-two magnetic unit cell via
        an angle-degeneracy parameter $\phi$. \textbf{(b)} Possible vectors
        $\vec M$ are shown as given in
            Eq.~\eqref{eq:order_parameter_original}. The black solid
            circle indicates $|\vec M|=1$, which is fulfilled for the
            ground state manifold depicted in (a). The arrows correspond to
            the four striped phases (light blue) and the
        four microvortex phases (dark blue). A pictograph is given to
        associate the vectors with their respective phases. The
    light blue dot in the middle corresponds to the paramagnetic phase.}
\end{figure}

    Irrespective of a truncation of the summation in
    Eq.~\eqref{eq:hamiltonian},
the ground state of the dXY system on the square lattice
is continuously degenerate and is
defined in the magnetic unit cell, which is a two-by-two
plaquette. The ground state spin configuration is parametrized by a global
angle-degeneracy parameter $\phi$, as depicted in
Fig.~\ref{fig:gs}~\cite{Belobrov1983}. This continuous degeneracy is broken by
finite temperature or dilution disorder as shown
in Ref.~\cite{Prakash1990}.
Namely, thermal
excitations favor striped phases, where $\phi=n\frac{\pi}{2}$ for $n\in
\mathbb Z$, due to different spin-wave stiffnesses along
different directions. In contrast to thermal excitations,
dilution disorder is known to select the so-called microvortex phase, where
$\phi=\frac{\pi}{4}+n\frac{\pi}{2}$ again with $n\in \mathbb Z$. The
microvortex phase ensures magnetic flux closure at the scale of each
plaquette, whereas magnetic flux closure happens in the striped phase at
infinity.

Up to now~\cite{DeBell1997,Carbognani2000,Rastelli2002,Fernandez2007,Baek2011},
any type of long range order in the dXY or the tdXY systems on the square
lattice has been described with the magnitude of the order parameter
\begin{align}
    |\vec M| &= \frac{1}{N}\left|\sum_i \left(
        \left( -1 \right)^{y_i}\cos\theta_i,
        \left( -1 \right)^{x_i}\sin\theta_i
    \right)\right|,
    \label{eq:order_parameter_original}
\end{align}
where $\theta_i$ is the angle of the $i$th spin with, for example, the
$x$-axis. The sites are enumerated with $x_i$ and $y_i$ along $\hat x$ and
$\hat y$. In the
non-disordered case, under the assumption of a nearest-neighbor distance of
$1$, the enumeration indices $x_i$ and $y_i$ are also the $x$ and $y$
coordinates respectively.
The order parameter is normalized to be $1$ for the ground states by dividing
by the total number of spins $N$.
The vector $\vec M$ lies on the unit circle for the ground state
configurations. Possible values for the vector $\vec M$ are depicted in
Fig.~\ref{fig:order-parameter-interpretation} and example ground states
are given as pictographs, which represent the character of the phase given by
each vector. As an example, the point $(1,0)$ corresponds to a striped order
along
$\hat x$, whereas $(-1,0)$ also corresponds to a striped order along $\hat x$
shifted by half a magnetic unit
cell along the $\hat y$ direction. Analogously, striped orders along $\hat y$
correspond to the two vectors $(0,\pm 1)$. The microvortex phases
correspond to the four points at $\frac{1}{\sqrt 2}(\pm 1,\pm 1)$ and the
paramagnetic phase corresponds to $(0,0)$.

Since the vector $\vec M$ lies on the circle described by $|\vec M|=1$ for all
ground state phases, it is not possible to distinguish the microvortex phase
from the striped phase by the magnitude $|\vec M|$. However, it is possible to
differentiate between the paramagnetic phase and long-range order in
either the microvortex phase or the striped phase.

In order to differentiate the ground state phases, we can consider
the
polar representation of the order parameter $\vec M=(M_x,M_y)=|\vec M|(\cos
\phi, \sin\phi)$.
The vector with doubled angle ($|\vec M|(\cos 2\phi,\sin 2\phi)$) is
introduced since this vector assigns the striped phases to vectors along the
$x$-axis and the microvortex phases to vectors along the $y$-axis.
This gives:
\begin{subequations}
    \begin{align}
        |\vec M|
          \cos\left( 2\arctan\left( \frac{M_y}{M_x} \right) \right)&=
          \frac{M_x^2-M_y^2}{|\vec M|},
          \label{eq:3a}
        \\
        |\vec M|
        \sin\left( 2\arctan\left( \frac{M_y}{M_x} \right) \right)&=
        \frac{2M_xM_y}{|\vec M|},
          \label{eq:3b}
    \end{align}
    \label{eq:double_angle_formulas}
\end{subequations}
which describe the projections of a state onto its striped phase
components and its microvortex phase components, respectively.

Eqs.~\eqref{eq:double_angle_formulas} therefore give possible order parameters
for (a) the striped and (b) the microvortex phase. These order parameters are,
however,
numerically unfavorable at high temperature, since
they are divided by the length of the vector $|\vec M|$. Group theory can
therefore be considered in order to find order
parameters with the same transformation properties. Such order parameters have
to transform as irreducible
representations of the symmetry group of the underlying system.
For the dXY model on the square lattice, the symmetry group is given by
time reversal symmetry enhanced by the corresponding point group of the
lattice, which is
$C_{4v}$ for the square lattice. The character table for this point group is
given in Table~\ref{tab:char_table_c4v}. In the last column of the table, the
simplest functions are indicated, which transform according to the
irreducible representations.
These functions are the symmetry-allowed combinations of
the components of the vector $\vec M$ used to construct the order parameters.

\begin{table}
    \centering
    \caption{Character table for $C_{4v}$, the point group of the square
    lattice.}
    \begin{tabular}{c|ccccc|c}
        $C_{4v}$& $E$& $2C_4$& $C_2$& $2\sigma_v$& $2\sigma_d$& $$\\\hline
        $A_1$& $1$& $1$& $1$& $1$& $1$& $x^2+y^2$\\
        $A_2$& $1$& $1$& $1$& $\overline 1$& $\overline 1$& $$\\
        $B_1$& $1$& $\overline 1$& $1$& $1$& $\overline 1$& $x^2-y^2$\\
        $B_2$& $1$& $\overline 1$& $1$& $\overline 1$& $1$& $xy$\\
        $E$& $2$& $0$& $\overline 2$& $0$& $0$& $(x,y)$\\
    \end{tabular}
    \label{tab:char_table_c4v}
\end{table}
The vector $\vec M$ itself transforms
according to the irreducible representation $E$, and therefore serves as an
order parameter. The length of the vector transforms according to the trivial
representation $A_1$ that, due to its transformation property, can only be used
to distinguish between long-range order and the paramagnetic phase. Inspection
of Table~\ref{tab:char_table_c4v} reveals that the two projections derived in
Eqs.~\eqref{eq:3a} and \eqref{eq:3b} transform according to the irreducible
representations $B_1$ and $B_2$, respectively, and therefore serve as valid
order parameters.

Thus, 
\begin{align}
    M_{{s }}=\sqrt{|M_x^2-M_y^2|}
    \quad\text{and}\quad
    M_{{mv}}=\sqrt{|2M_xM_y|}
    \label{eq:mc_order_params}
\end{align}
are also valid order parameters for the striped phase and the microvortex
phase, since they transform according to $B_1$ and $B_2$
respectively. Furthermore, $M_s$ and $M_{mv}$ are numerically more stable as
they do not contain a division by the
magnitude $|\vec M|$. These two quantities as well as $|\vec M|$ are determined
in the subsequent Monte Carlo simulations in order to distinguish between the
different phases.
\section{Monte Carlo Simulations}
\label{sec:mc}
Monte Carlo simulations are now performed for the tdXY system on the
square lattice in order to construct the phase diagrams for both the
dilution-disordered system as well as the random-displacement disordered
system. The code \footnote{Source code available under:
\url{http://github.com/domischi/mcpp}} is based on the ALPS
project~\cite{Troyer1998,Albuquerque2007,Bauer2011}. It uses a parallel
tempering algorithm~\cite{Swendsen1986,Hukushima1996,Katzgraber2006} (also
known as replica-exchange Monte Carlo) in order to thermalize even quite
heavily frustrated systems. Parallel tempering refers to the simulation of the
same system at several temperatures in parallel, with regular exchange of the
temperatures between the simulations
according to a detailed-balance condition.
All figures are generated with matplotlib~\cite{Hunter2007}.

In the following simulations periodic boundary conditions were
used. We thermalized the system with
$2\cdot 10^5$ lattice sweeps. Subsequently, $10^4$ measurements were made
while, between two successive measurements, 15 lattice sweeps were carried
out.
For the disordered cases, a total of 40 different temperatures were used,
which were linearly distributed between $T=0.1$ and $T=1.6$. The
simulations performed for this work took a total of approximately $10^5$
CPU hours, with the
majority of time spend on simulating the disordered systems, where the
disorder average over several realizations had to be taken.
Before considering the simulations for the disordered systems
the non-disordered case is first simulated to gain
further insight into
this simpler situation where no disorder average has to be
taken.

\subsection{No Disorder}
\label{sec:non-disordered}
\begin{figure}
    \subfloat[][]{
        \includegraphics[width=0.8\columnwidth]{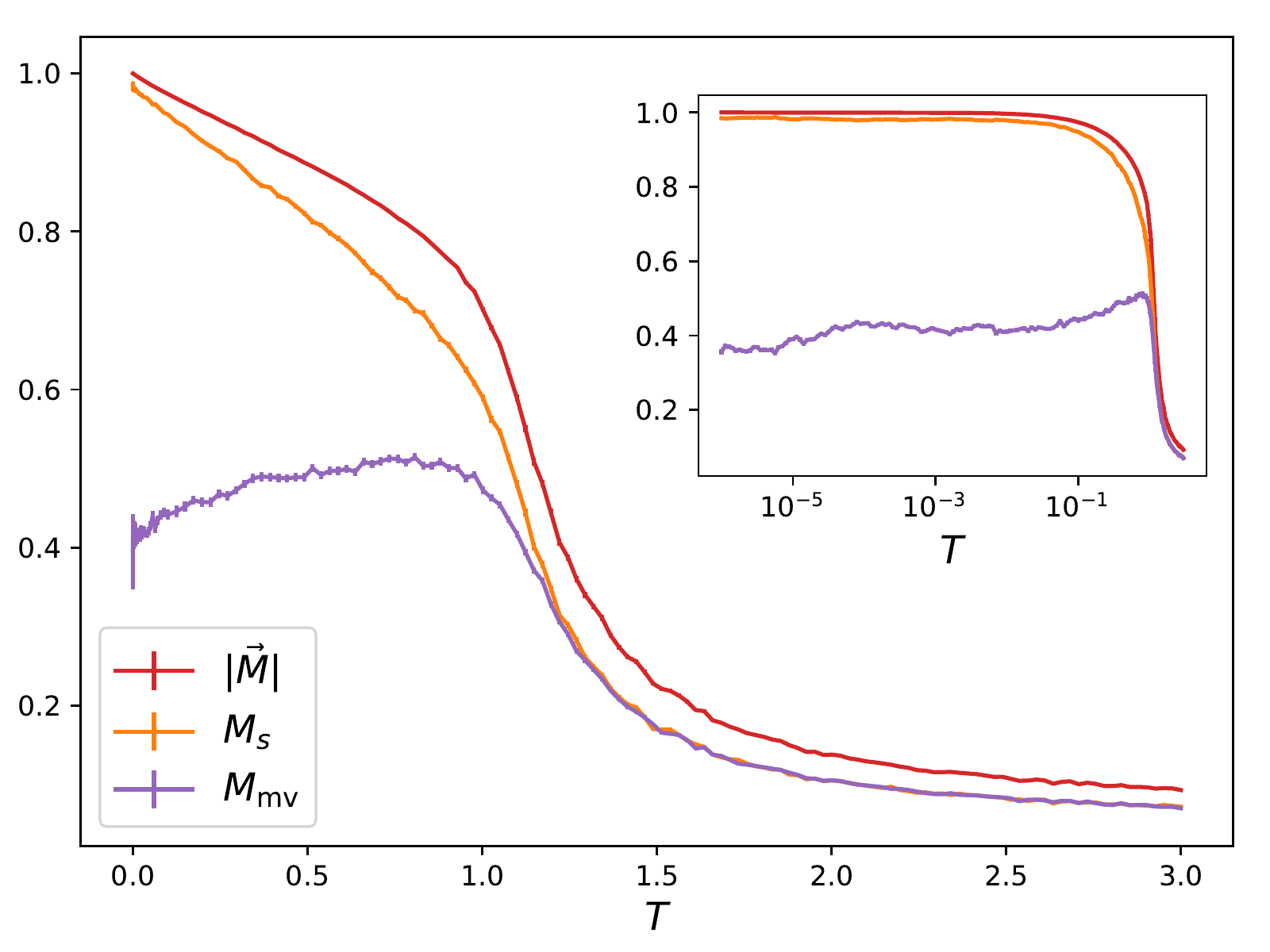}
        \label{fig:non-disordered}
    } \
    \subfloat[][]{
        \includegraphics[width=0.8\columnwidth]{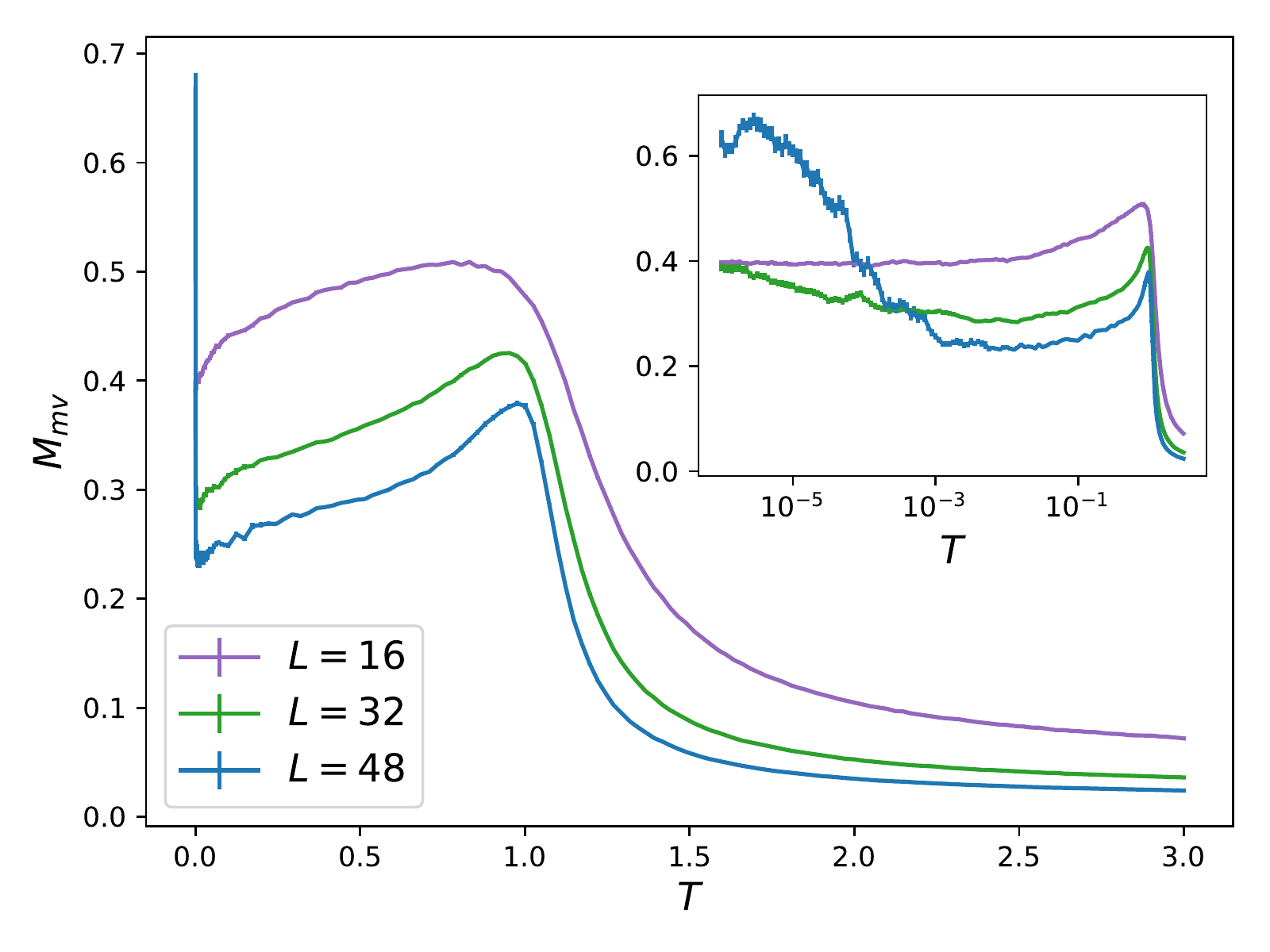}
        \label{fig:T0-PT}
    } \
    \subfloat[][]{
        \includegraphics[width=.7\columnwidth,height=.7\columnwidth]
        {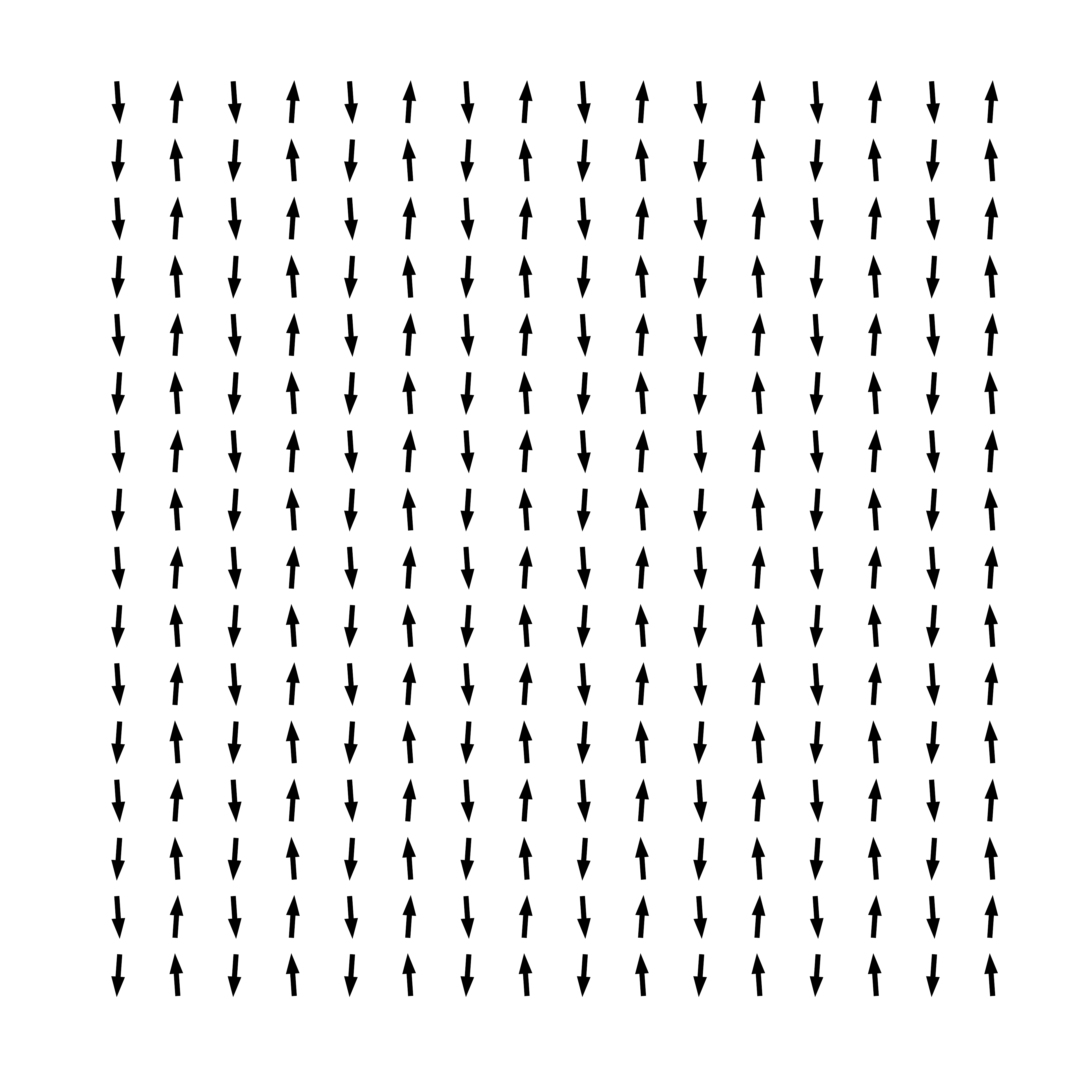}
        \label{fig:non-disordered-config}
    }
    \caption{(color online)
        \textbf{(a)} The temperature dependence of the three order
        parameters discussed in section~\ref{sec:derivation_order_parameter}
        are shown. These are obtained with parallel tempering Monte
        Carlo simulations for the non-disordered tdXY system with a
        system size of
        $L=16$ averaged over $4$ independent runs.
        The same data plotted on a logarithmic temperature scale is
        shown in the inset.
            \textbf{(b)} The microvortex order parameter is shown for the three
            system sizes studied. The same data is plotted on a logarithmic
            temperature scale in the inset to highlight the transition at
            $T=0$.
        \textbf{(c)}
        A configuration is shown, which was obtained by our simulations at
        temperature $1.8\cdot 10^{-6}$. Since the microvortex order
        parameter is very homogeneous across the system, it is likely to
        originate from the Goldstone mode.
    }
\end{figure}
The simulation for the non-disordered system was performed in order to
validate the order parameters derived in
section~\ref{sec:derivation_order_parameter}. In total, 220
temperatures were implemented, which were
uniformly spaced at higher temperatures and logarithmically spaced at lower
temperatures.
The three order parameters are plotted
versus the temperature in Fig.~\ref{fig:non-disordered}.
As expected, the magnitude $|\vec M|$ indicates the appearance of long-range
order as the temperature decreases (obtained from the Binder cumulant crossing:
$T_c(\rcut=2)=0.996\pm 0.017$). A similar trend is visible for $M_s$.
Furthermore, fluctuations away from the striped phase have a contribution
    to
    $M_{mv}$, so that $M_{mv}$ rises around the phase transition between the
paramagnetic phase and the striped phase and then slowly decays. 

In the
inset of Fig.~{\ref{fig:non-disordered}}, we show the same data with a
logarithmic temperature axis. 
    The
    value of $M_{mv}$ appears to saturate at around $0.4$. Comparing, however,
    with Fig.~\ref{fig:T0-PT}, where the data for $M_{mv}$ is shown
    for three different system sizes it can be seen that there is a steep
    increase of $M_{mv}$, which indicates the transition to the angle
    degenerate ground state~\cite{Prakash1990}.
    This transition is driven by the Goldstone mode, which is a result of the
    angle degenerate ground state, transforming in this case
    $\phi\rightarrow \phi+\delta\phi$ in Fig.~\ref{fig:gs} over a large length
    scale. This Goldstone mode can be seen in
    Fig.~\ref{fig:non-disordered-config}, where $M_{mv}$ is distributed very
    homogeneously across the system.
    Note that the Goldstone mode makes all angles $\phi$ in Fig.~\ref{fig:gs}
    equally accessible, so that the saturation value for $M_{mv}$ in the limit
    $T\rightarrow 0$ can be computed as an average with respect to $\phi$.
    Doing so gives $M_{mv}\approx 0.7628$ for $T\rightarrow 0$, which is
    consistent with the trend
    seen in the inset of Fig.~\ref{fig:T0-PT} for the largest system size
considered.

To conclude the results of this section, the order parameters $M_s$ and
$M_{mv}$ introduced in Eq.~\eqref{eq:mc_order_params} give a measure of the
striped phase and the microvortex phase respectively. Therefore they can be
used in the subsequent
simulations in the two disordered cases to obtain the phase diagrams
of the tdXY systems.

\subsection{Dilution}
\label{sec:dilution}
We now consider dilution disorder through the introduction of vacancies.
Starting with the non-disordered square lattice, moments are removed with a
probability $p$, which will be referred to as the dilution rate.

The diluted square-lattice dXY was previously treated
using a spin-wave calculation in order to obtain the phase
diagram~\cite{Prakash1990}. In this spin-wave calculation a truncation of
$\rcut=1$ was applied. For small but finite $p$ at $T=0$, the
microvortex phase is preferred
and for small
but finite $T$ at $p=0$, the striped phase is preferred.
For any value of $p$ at sufficiently high $T$, the paramagnetic phase is
expected.
With temperature-sweep Monte
Carlo simulations, a first quantitative phase
diagram was constructed~\cite{DeBell1997,Patchedjiev2005,LeBlanc2006}.
Here, the measured observable consisted of fourth powers of the spin
components and essentially was a measure of the likelihood that spins point
along diagonals rather than along axes.
This gave an indication of the
selected phase, but did not serve as an order parameter. This
led to well-converged results for small values of $p$.
However, due to the frustration and the disorder at higher
values of the dilution rate, a temperature-sweep algorithm
is prone to get stuck in metastable states, so that no
conclusive statement was possible above a dilution rate of approximately
$6\%$.

\begin{figure}
    \centering
    \includegraphics[width=\columnwidth]{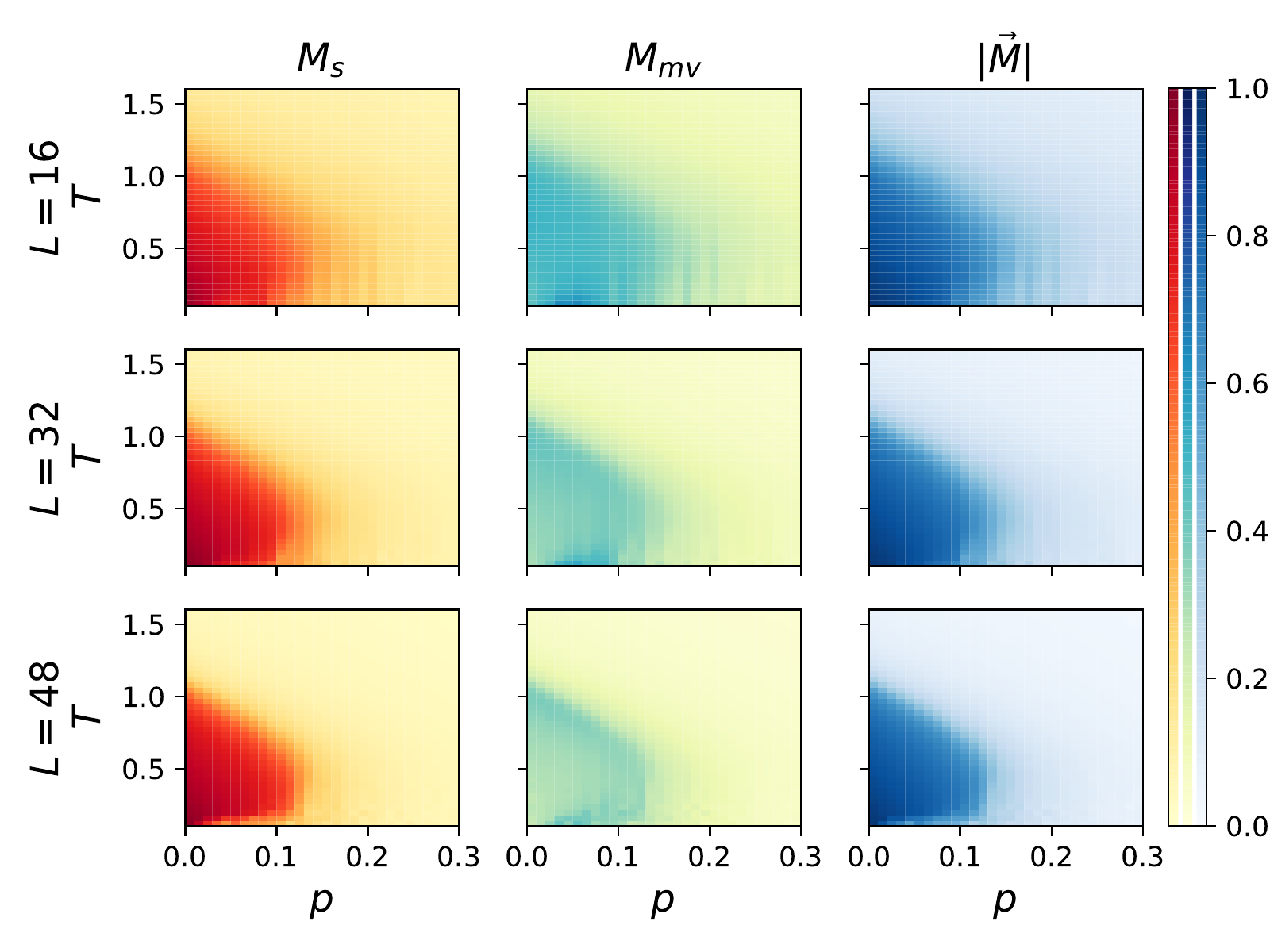}
    \caption{(color online) Each of the order parameters for the
        dilution-disordered tdXY system on the square lattice as a
        function of dilution rate $p$ and temperature $T$ for three different
        system sizes ($L=16,32,48$) obtained via parallel
        tempering Monte Carlo
        simulations averaged over 32
        disorder realizations.}
        \label{fig:dil}
\end{figure}

The order parameters determined by our Monte Carlo simulations as a function
of temperature and dilution rate are summarized in Fig.~\ref{fig:dil}. There is
good convergence of the data
for all system sizes, temperatures and dilution rates since there is no visible
noise.
The previously proposed phase diagram~\cite{Prakash1990} is
in qualitative agreement with
the results for $M_s$ and $M_{mv}$.
Namely, there is a pocket at low
temperatures and finite dilution rate where the microvortex phase is
predominant (region with strong signal in the panels for $M_{mv}$, which is
more visible at larger system sizes). In
addition, for small dilution rates and high enough temperatures, the striped
phase dominates (region with a strong signal in the panels for $M_s$).
Nonetheless, in regions where one phase dominates, there is still some
signal of the order parameter for the other phase visible. This occurs because
fluctuations from one phase appear as an increase in the
order parameter of the other phase.

Previously, it was predicted for a nearest-neighbor truncated
dipolar Hamiltonian~\cite{Prakash1990} that any long-range order
disappears close to
the percolation limit of the square lattice at
$1-p_{c}^{\text{perc}}=40.7\%$. This is in agreement with the
general expectation for nearest-neighbor only Hamiltonians,
that no long-range order can be seen above the percolation
threshold. 
    However, inspection of $|\vec M|$ in
    Fig.~{\ref{fig:dil}}
reveals that there is no longer a sizable contribution to the long-range
order parameter already at a dilution rate of $p_c(\rcut=2)\approx
15\%$. Note that this value is dependent on the cutoff and that the
inclusion of more lattice
sites leads to a reduction of both $T_c$ as well as $p_c$ due to the
increase in frustration present in the system. This is in contrast to
percolation theory,
which predicts an increase of $p_c$ as $\rcut$ is increased.

All of the data presented in Fig.~\ref{fig:dil} is system-size dependent.
In order to give a system-size independent phase diagram another method
needs to be implemented. Even though the data is not good enough to attempt a
scaling collapse, a Binder cumulant analysis can be applied. Crossings of the
cumulants for different system sizes at a fixed dilution rate can, up to
corrections to scaling, precisely locate the
critical temperatures for the involved
transitions.

Making use of the binning analysis implemented in ALPS, we obtain the Binder
cumulants
with their statistical error.
Through a resampling
procedure, such error information can be used to obtain
possible realizations of the Binder cumulant curves. In particular, the mean
value of the Binder cumulants as a function of $T$ at a fixed value of $p$ was
perturbed
with uncorrelated Gaussian noise according to
the statistical error at each sampling point.
Through the analysis of many such curves, statistics on the crossings can
be obtained and, from this, an estimate for $T_c$ and its uncertainty at
every value of $p$ can be determined. We refer to this method as the
\textit{fixed dilution rate
analysis}. Analogously, the same procedure can be applied for the
Binder cumulants at a
fixed temperature as a function of $p$ in order to obtain an estimate for $p_c$
at every value of $T$. We refer to this as the \textit{fixed temperature
analysis}.

\begin{figure}
    \centering
    \includegraphics[width=\columnwidth]{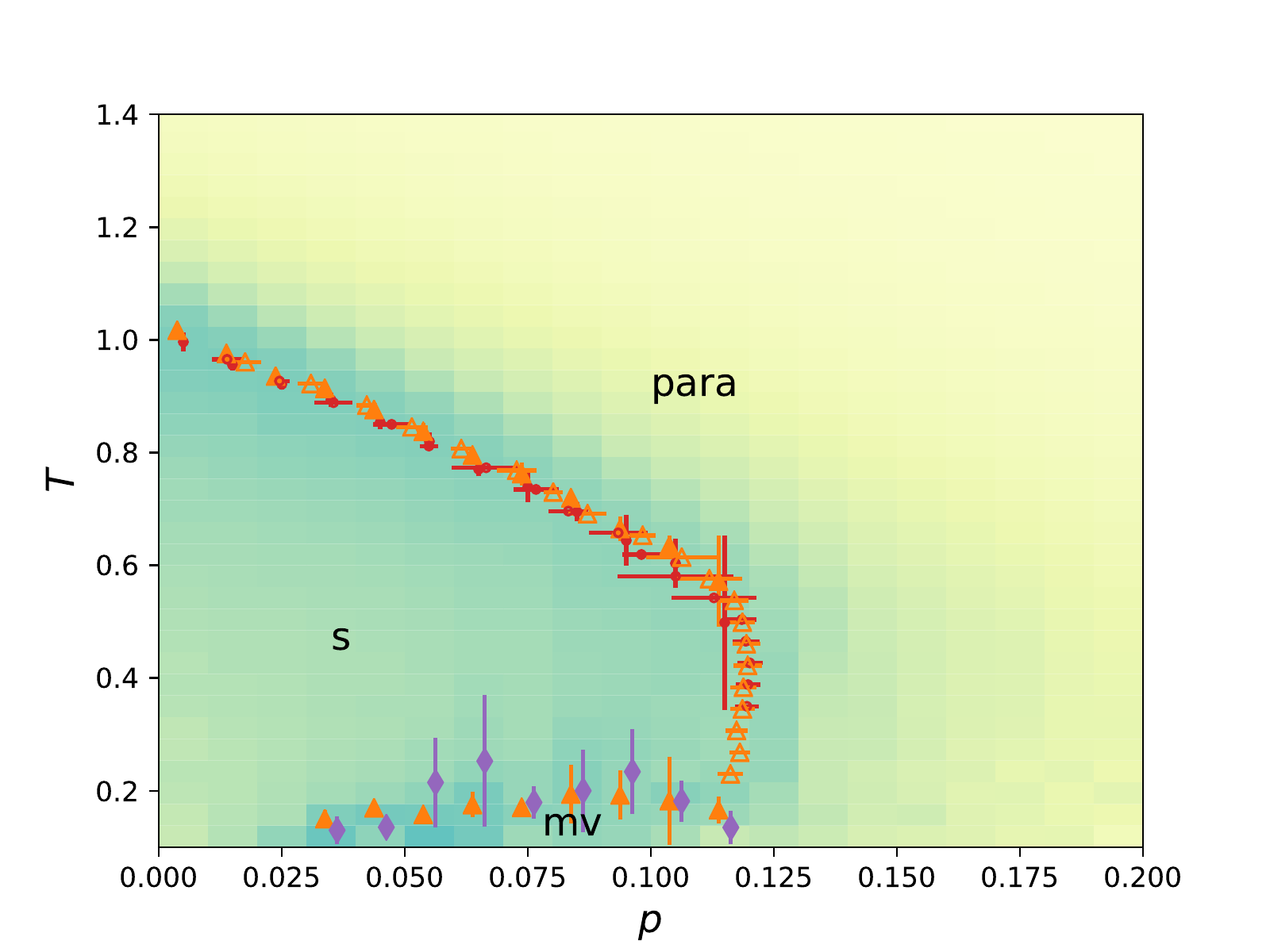}
    \caption{(color online)
        The phase diagram for the diluted square lattice tdXY system as a
        function of dilution rate and temperature derived via Binder cumulant
        crossings,
        superimposed on the corresponding $M_{mv}$ data of
        Fig.~\ref{fig:dil} for $L=48$. The filled (open) markers give $T_c$
        ($p_c$) with red dots for
        $|\vec M|$, orange triangles for $M_s$ and violet diamonds for $M_{mv}$
        respectively. Region ``mv'' corresponds to the microvortex
        phase, region ``s'' to the striped phase and region ``para'' to the
        paramagnetic phase.
    }
    \label{fig:dil-pd-48}
\end{figure}

The system-size independent phase diagram is shown in
Fig.~\ref{fig:dil-pd-48}. Filled markers denote
the procedure where the data was analyzed for a fixed dilution rate to obtain
$T_c$, whereas
open markers are the data for $p_c$ obtained with the fixed temperature
analysis.
The Binder cumulant estimate for $T_c$ ($p_c$) are shown with red dots, violet
diamonds and orange triangles for $|\vec M|$, $M_{mv}$ and $M_{s}$,
respectively.
For comparison, the microvorticity heat map ($M_{mv}$) for $L=48$ is shown in
the
background.

A few remarkable features can be identified in Fig.~\ref{fig:dil-pd-48}. At the
critical line separating the paramagnetic phase and the striped phase, $T_c$
predicted by the Binder cumulant analysis of $M_s$ and $|\vec M|$ agree well
and have small error bars. This data also agrees well with the data for
$p_c$, which was obtained by the fixed temperature analysis. Furthermore, the
fixed
temperature analysis yielded the boundary between the striped phase and the
paramagnetic phase at
$p_c(T=0, \rcut=2)\approx 11\%$, a value which is system size independent,
    in
contrast to the earlier estimate of 15\%.
    Note that $p_c$ is cutoff dependent. 
    Since a larger \rcut{} increases the frustration, we expect that 
    $p_c(T=0,\rcut=\infty)\leq p_c(T=0, \rcut=2)$, so that our result for $p_c$
    serves as an upper bound to $p_c(\rcut=\infty)$. In fact, the frustration
    could even lead to $p_c(\rcut\to\infty)\to 0$.

At $p>11\%$,
our fixed dilution rate analysis could no longer provide quantitative data.
This is due
to the fact that the phase boundary is close to vertical, so that $T_c$ in this
area is very sensitive to $p$.
At the lower critical line, separating the microvortex
phase (region ``mv'') from the striped phase (region ``s''), there is again
good
agreement between the data for the Binder cumulants of $M_{mv}$ and $M_s$.
However, here the error bars
are substantially larger. This
is due to the Binder cumulants
being flatter as a function of temperature,
resulting in poorly defined crossings.
The fixed temperature analysis did not perform well for the phase boundary
between the microvortex phase and the striped phase since it was noise
dominated. Therefore this data is not shown.
At $p\geq 12\%$, the analysis could no
longer be performed, as there were no more crossings. This corresponds to the
onset of paramagnetism in the region labeled ``para''.

\subsection{Random displacement}
\label{sec:random_displacement}
We now introduce random displacement.
Starting with
the non-disordered square lattice, every site is relocated by a random
displacement in the $xy$-plane, taken from a Gaussian random distribution with
standard deviation $\sigma$.

To the best of our knowledge, no attempt has been made to provide a phase
diagram with respect to the strength of the random displacement,
even though simulations have been performed for both medium
disorder~\cite{Alonso2011} and strong disorder given by
random placement of the moments~\cite{Pastor2008}.
For the work on medium disorder~\cite{Alonso2011} the starting point was the
square lattice and random displacements taken from a Gaussian
distribution were introduced. This paper
was mainly concerned with the disappearance of long-range
order with the appearance of a possible spin glass phase for higher
values of disorder. The random placement of moments~\cite{Pastor2008} resulted
in a spatial
localization of magnetic excitations as well as low-energy states
incorporating microvortex-like structures. The scope of this
work was, however, mainly the low-energy excitations and not
the construction of the phase diagram.

\begin{figure}
    \subfloat[][]{
        \includegraphics[width=\columnwidth]{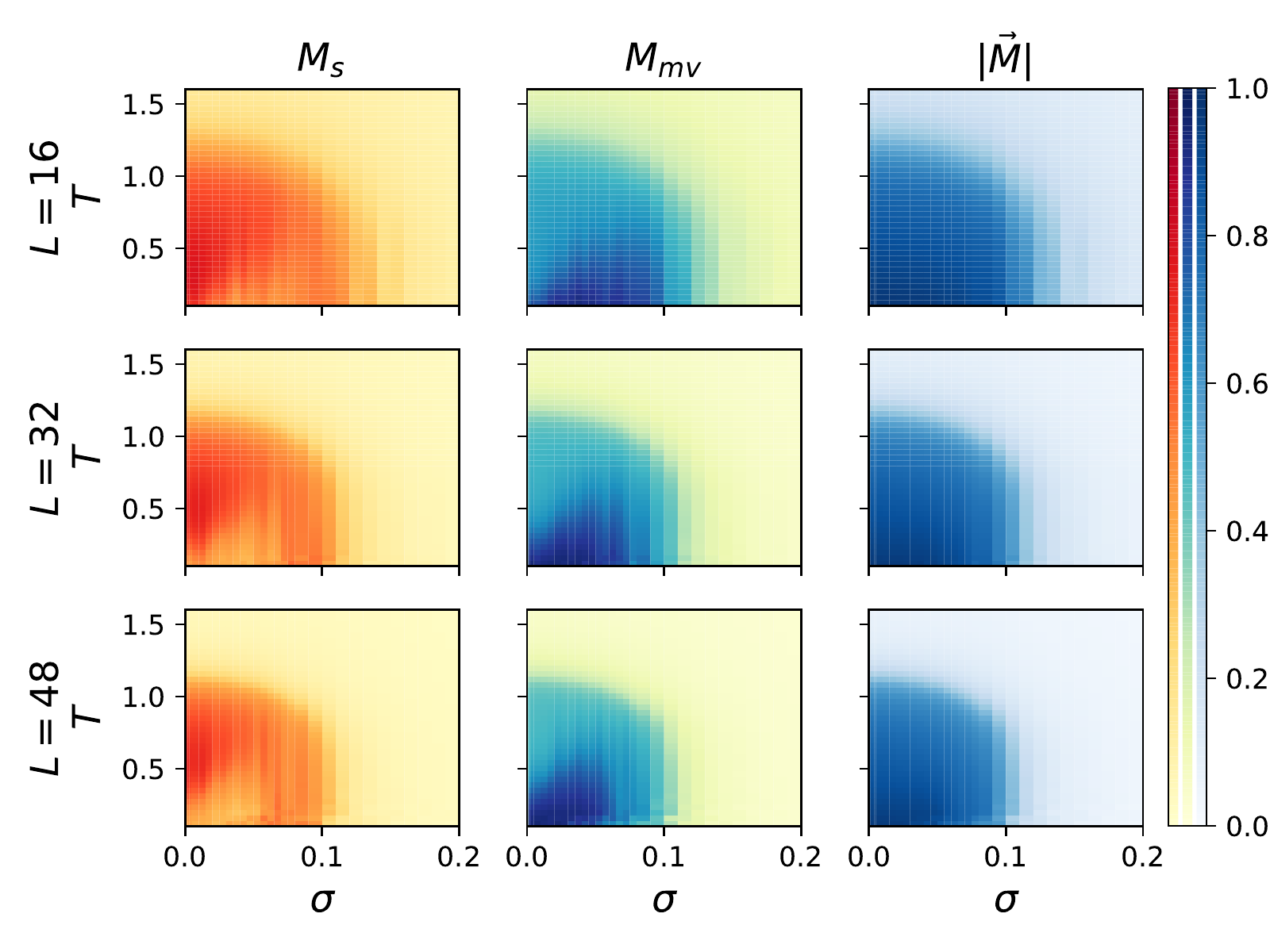}
        \label{fig:pos}
    } \
    \subfloat[][]{
        \includegraphics[width=\columnwidth]
        {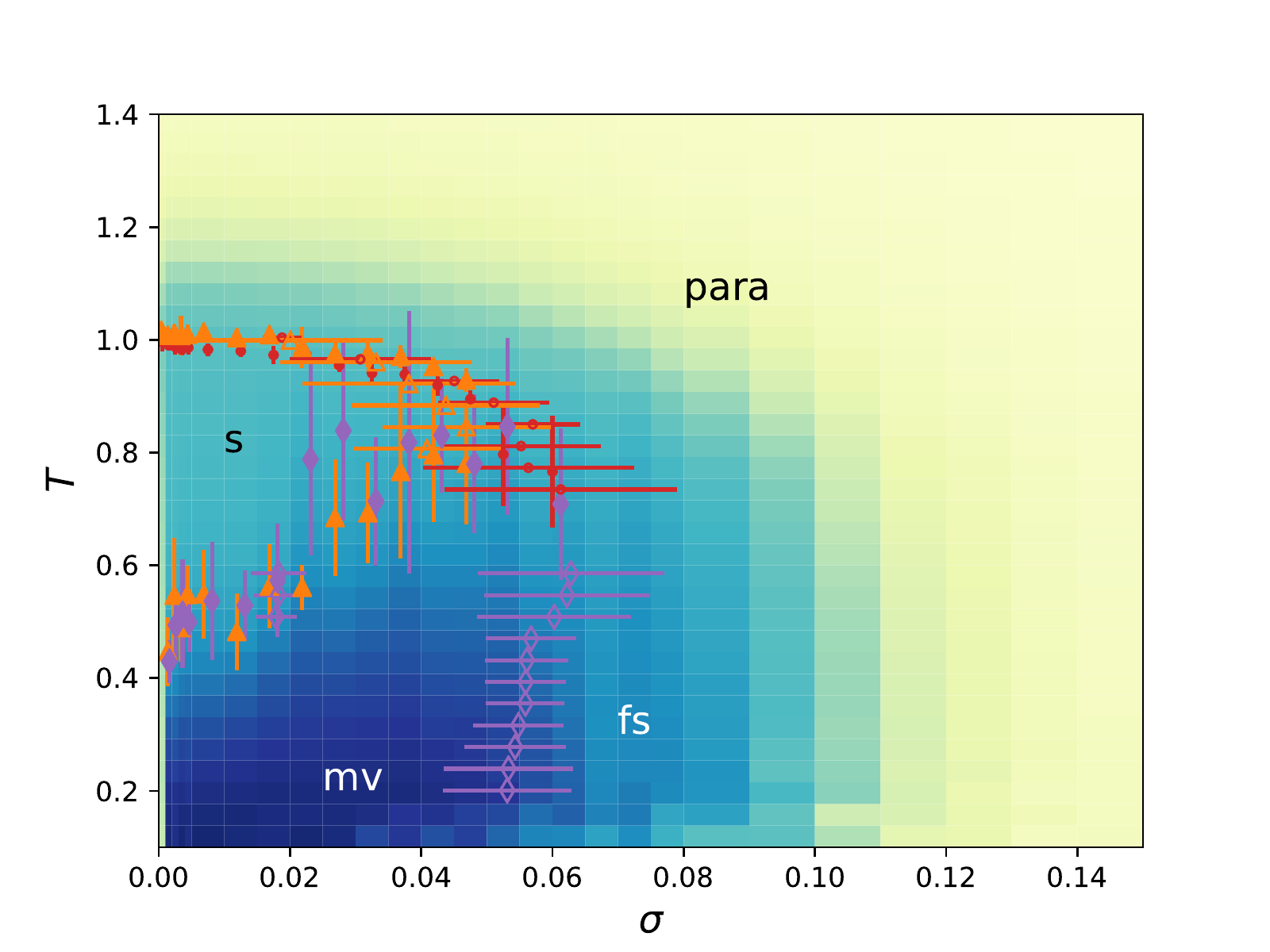}
        \label{fig:pos-pd-48}
    }
    \caption{(color online)
        \textbf{(a)} Each of the order parameters for random-displacement
        disordered tdXY systems
        on the square lattice as a function of the width of the random
                displacement $\sigma$ and temperature $T$ for
        three different system sizes ($L=16,32,48$)
        obtained via parallel tempering Monte Carlo simulations averaged over
        32 realizations.
        \textbf{(b)}
        The phase diagram of the same system as (a), as a function of
        temperature and
        width of the random displacement. This is derived via Binder
        cumulant crossing and superimposed on the
        corresponding $M_{mv}$ data for $L=48$ of (a).
        The assignment of colors, markers and regions is the same as in
        Fig.~\ref{fig:dil-pd-48}. The region ``fs'' not present in the dilution
        case denotes region dominated by finite size effects.}
\end{figure}

Our results for the three order parameters are shown in
Fig.~\ref{fig:pos}. Similar to the dilution-disordered case,
for random-displacement disorder, the system also favors the microvortex phase
at low
temperatures with a pocket of large values of the microvortex order
parameter for small $T$ and intermediate $\sigma$. Likewise, a high enough
temperature results in the striped phase (small $\sigma$ and intermediate $T$).
Analogous to the dilution-disordered case presented in
section~\ref{sec:dilution}, the heat maps for $|\vec M|$ indicate where no
sizable contribution to long-range order is expected, which
occurs at approximately $\sigma>0.12$.

While the phase diagrams in Fig.~\ref{fig:dil-pd-48} and \ref{fig:pos-pd-48}
are for two different types of disorder, they
    should agree at $p=0$ and $\sigma=0$, since here both simulations are
    non-disordered. At first glance, this does not appear to be the case,
    but looking closely at the phase diagram of the random displacement
    disordered system at small $\sigma$, the data does indeed
    agree with the non-disordered system and behaves continuously with
    $\sigma$. However, even small values of $\sigma\lesssim0.01$ are
    sufficient to stabilize the microvortex phase up to considerably high
    temperatures, which explains the apparent mismatch between the two
figures.

As carried out for the dilution-disordered system, a Binder cumulant analysis
was performed for the random-displacement disordered system. The results are
displayed in Fig.~\ref{fig:pos-pd-48}, where again the
microvorticity order parameter for $L=48$ is displayed in
the background to serve as a reference. The definitions of the marker
colors, forms and fillings for the data points of the Binder cumulant
crossings are the same as for the dilution-disordered case.
Again, the regions in the figure correspond to the
microvortex phase (mv), the striped phase (s) and the
paramagnetic phase (para). At some phase boundaries it was not possible to
determine the Binder cumulant crossing for the fixed
temperature analysis due to statistical noise and
data points are only shown where the analysis could reliably be performed.
Once more, a good agreement between the data for the different order
parameters can be seen. The Binder cumulant analysis at a fixed disorder
strength breaks down at a disorder strength
of $\sigma_c(T=0,\rcut=2)\approx 0.06$. This is again cutoff
dependent. As argued in the dilution disordered case, we expect analogously
$\sigma_c(T=0,\rcut=\infty)\leq \sigma_c(T=0,\rcut=2)$.
Note that non-negligible values of the
microvortex parameter persist up to $\sigma\approx 0.11$, which is much
larger
than $\sigma_c(T=0,\rcut=2)$. The appearance of the associated region
``fs'' can be explained by the finite size of the simulations.
Indeed, this region becomes smaller as the system size gets bigger
as
seen in Fig.~\ref{fig:pos}, middle column.

\section{Applicability of the order parameters}
\label{sec:applicability}
Strictly speaking, the order parameters ($|\vec M|$, $M_s$, $M_{mv}$) are only
valid for
the non-disordered system, since any disorder will in principle invalidate the
symmetry discussion made in section~\ref{sec:derivation_order_parameter}.
Nevertheless, for small disorder, the derived order parameters should still be
approximately valid. The implicit assumption made to employ the derived order
parameters, even in disordered systems, is that the
enumeration indices $x_i$ and $y_i$ in
Eq.~\eqref{eq:order_parameter_original} are approximately valid
descriptions of the lattice positions. Certainly for
small disorder the indices specify the positions well. However,
in the highly disordered systems, this is no longer true.

To test if the enumeration in terms of $x_i$ and $y_i$ is valid,
the random-displacement disordered system can be considered.
The problem of enumeration becomes apparent when two moments exchange their
relative order. This is formally written as follows: let us denote the position
of the $i$th
moment in the non-disordered case with $\vec r=(r^x_i,r^y_i)$ and the position
after applying the disorder with $\vec R=(R^x_i,R^y_i)$.
To compute the exchange probability, consider now two sites
$i$ and $j$, which respect in the non-disordered case $r^x_i<r^x_{j}$. An
exchange along the $x$-direction has occurred if $R^x_i>R^x_{j}$. Analogously
for the $y$-direction,
$r^y_i<r^y_{j}$ but $R^y_i>R^y_{j}$.
The probability of an exchange event depends on the width of the random
displacement and can be computed to be
\begin{align}
    \label{eq:probability_breakdown}
    \rho_{\text{ex}}(\sigma)&=
    2\cdot\frac{1}{2\pi\sigma^2}
    \int_{-\infty}^{\infty}\mathrm{d}v
    \int_{v}^{\infty}     \mathrm{d}u\:
    e^{-\frac{u^2}{2\sigma^2}     }
    e^{-\frac{(v-1)^2}{2\sigma^2} }.
\end{align}
The factor $2$ comes from considering the exchange of sites along both the
$x$-direction as well as the $y$-direction.

As soon as an exchange event occurs, the group theoretical symmetry
discussion in section~\ref{sec:derivation_order_parameter}
will be invalidated.
Therefore we need to make sure that the value of $\rho_{\text{ex}}(\sigma)$ is
small enough, so that the order parameters obtain from the simulations are well
defined in each
region of the phase diagram.
For
example, the exchange probability given in Eq.~\eqref{eq:probability_breakdown}
can be computed for the largest $\sigma$ used in
section~\ref{sec:random_displacement}, namely a standard deviation of
$\sigma=0.2$, resulting in $\rho_{\text{ex}}(\sigma=0.2)=4\cdot
10^{-4}$. This
exchange probability appears to be relatively high considering that, even in
the $L=16$
system, the exchange of two sites is expected to happen a total of $3$ times in
$32$ disorder realizations. Therefore in approximately $10\%$ of the
simulations, the definition of the order parameter breaks down at least
locally.
However, this is the highest disorder considered and the system is already in
the paramagnetic phase, so that an error of this size should not affect the
conclusions considering the phase diagram.

For smaller values of $\sigma$, the exchange probability diminishes
drastically. To
illustrate this, for $\sigma=0.16$, which is just slightly smaller than
the highest disorder considered, the exchange is expected to only occur once in
the 32 disorder realizations for the largest considered system-size ($L=48$)
and
this is still deep in the paramagnetic phase. Below $\sigma=0.16$,
$\rho_{\text{ex}}\approx 0$ so it is not
expected that the order parameters break down at all in the simulations for low
$\sigma$.
Therefore, we can conclude that the order parameter definitions in
Eq.~\eqref{eq:mc_order_params} are well
justified for the construction of the phase diagrams.

\section{Conclusions}
\label{sec:conclusion}
In this work a truncated version of dipolar-coupled XY (tdXY) spin
system on the square lattice was treated under the
influence of disorder with Monte Carlo simulations.
Starting from the perfect lattice, disorder was introduced in two different
forms, namely by introduction of vacancies and by random displacement of each
site.
Some features of these systems are already known from previous
work~\cite{Belobrov1983,Prakash1990,DeBell1997,Patchedjiev2005,LeBlanc2006,
Pastor2008,Alonso2011,Baek2011}.
This paper extends these results by first deriving order parameters
for the phases known as the striped phase and the
microvortex phase. The order parameters of the tdXY system were then
determined using parallel
tempering Monte Carlo simulations, first for non-disordered systems down to
very low temperatures, and then for systems with either dilution or random
displacement as sources of disorder.
The phase diagrams for both cases of disorder were
obtained via a Binder cumulant approach, to find system-size independent values
for $T_c$ ($p_c$, $\sigma_c$) as well as to quantify
the uncertainty. Finally
it was argued that the definitions of the order parameters are well defined
even in the disordered systems.

The newly derived order parameters, as
well as the use of parallel tempering Monte Carlo simulations, allowed us to
distinguish the long-range ordered phases from the
paramagnetic phase, and to determine the character of the
long-range ordered phases. For both types of disorder,
well-converged results for the order parameters as a function of temperature
and disorder strength were obtained for all system sizes.
Furthermore, the system-size independent phase diagram could be derived via
a Binder cumulant analysis, which gave in most regions small error bars for
$T_c$ as well as $p_c$ and $\sigma_c$.

In previous work on the dilution-disordered
system~\cite{Prakash1990,DeBell1997,Patchedjiev2005,LeBlanc2006}, it was
speculated that the
disappearance of long-range order would occur close to the percolation
threshold of the square lattice ($1-p_{c}^{\text{perc}}=40.7\%$). In contrast
to these predictions, our simulation result for $\rcut=2$ showed a much lower
critical dilution rate of $p_c(T=0, \rcut=2)\approx 11\%$.
This value should serve at least as an upper
    bound on $p_c(\rcut=\infty)$.

    A full phase diagram for the random-displacement
    disordered
    tdXY system on the square lattice was obtained. In contrast to the
    dilution-disordered system, a large region seemed to be
    apparent (region ``fs'' in Fig.~\ref{fig:pos-pd-48}), where the
    system size results in a sizable contribution to the order parameters. This
    region ``fs'' is expected to vanish in the thermodynamic limit.
    Also for this system an upper bound on the critical disorder strength
        of $\sigma_c(\rcut=2)\approx 0.06$
    was obtained, above which no long-range
    order is expected.

    Interestingly, in the phase
    diagrams for both the dilution-disordered system as well as the
    random-displacement disordered system, the other regions behave similarly,
    even though the notion of disorder in the two systems is quite different.
    In particular, the microvortex
    phase is favored by both types
    of disorder. Also, in both systems, at high enough temperature
    and small enough disorder, the striped phase is
    favored, before ending in the paramagnetic phase at higher
    temperatures or disorder strengths.

    These similarities in the phase diagrams suggest a
    general mechanism for the selection of the microvortex phase, which
    is common to both dilution and random displacement. This can be understood
    by considering the fact that in disordered systems, in contrast to the
    non-disordered system, it is
    more difficult for magnetic flux closure to
    occur at bulk
    length scales, since disorder breaks locally many of the previously
    available symmetries. Therefore, instead of a global magnetic flux closure
    obtained by the
    striped phase, a more local magnetic flux closure structure as in the
    microvortex phase is likely to be
    favorable. The derivation of this more general mechanism from an analytic
    perspective poses an interesting question for future work.

    The data is openly available at \url{http://doi.org/10.5281/zenodo.1326251}
\begin{acknowledgments}
    We would like to thank Michael Sch\"utt for helpful discussions. We also
    would like to
    thank Na\"emi Leo and Valerio Scagnoli for carefully
    reading the manuscript and providing useful suggestions.
    This work was partially funded via a PSI-CROSS proposal (no. 03.15).
\end{acknowledgments}

\bibliography{library}
\end{document}